\pdfoutput=1
\documentclass[aps,10pt,twocolumn,pre,byrevtex,bibnotes,superscriptaddress,footinbib,showpacs,notitlepage,showkeys,longbibliography]{revtex4-1}
\usepackage{microtype}
\usepackage{graphicx}
\usepackage{amsmath}
\usepackage{amssymb}
\usepackage{amsthm}
\usepackage{color}
\definecolor{myblue}{rgb}{0.153,0.322,0.706}
\usepackage[colorlinks,linkcolor=myblue,urlcolor=myblue,citecolor=myblue,breaklinks=true]{hyperref}

\newcommand{\be}{\begin{equation}}
\newcommand{\ee}{\end{equation}}
\newcommand{\ra}{\rightarrow}
\newcommand{\p}{\partial}

\newcommand{\tI}{\bar I}
\newcommand{\reals}{\mathbb{R}}

\newcommand{\bx}{\ensuremath{\bar x}}
\newcommand{\bz}{\ensuremath{\bar z}}
\newcommand{\bv}{\ensuremath{\bar v}}
\newcommand{\bb}{\ensuremath{\bar\beta}}

\newcommand{\dx}{\ensuremath{\dot x}}

\newcommand{\ddx}{\ensuremath{\ddot x}}
\newcommand{\mr}[1]{\mathrm{#1}}
\newcommand{\s}{\sigma}
\newcommand{\al}{\alpha}
\newcommand{\g}{\gamma}
\newcommand{\e}{\epsilon}
\newcommand{\D}{\mathcal{D}}
\newcommand{\La}{\mathcal{L}}
\newcommand{\Sa}{\mathcal{S}}

\newcommand{\OO}{\mathcal{O}}
\newcommand{\bI}{\bar I}

\graphicspath{{newfigs/}}

\begin{document}
\title{\vspace*{0.06in} Noise correction of large deviations with anomalous scaling}

\author{Daniel Nickelsen}
\email{danieln@aims.ac.za}
\affiliation{African Institute for Mathematical Sciences (AIMS), Muizenberg 7950, South Africa}

\author{Hugo Touchette}
\email{htouchette@sun.ac.za, htouchet@alum.mit.edu}
\affiliation{Department of Mathematical Sciences, Stellenbosch University, Stellenbosch 7600, South Africa}

\date{\today}

\begin{abstract}
We present a path integral calculation of the probability distribution associated with the time-integrated moments of the Ornstein--Uhlenbeck process that includes the Gaussian prefactor in addition to the dominant path or instanton term obtained in the low-noise limit. The instanton term was obtained recently [D.~Nickelsen, H.~Touchette, Phys.\ Rev.\ Lett.\ \textbf{121}, 090602 (2018)] and shows that the large deviations of the time-integrated moments are anomalous in the sense that the logarithm of their distribution scales nonlinearly with the integration time. The Gaussian prefactor gives a correction to the low-noise approximation and leads us to define an instanton variance giving some insights as to how anomalous large deviations are created in time. The results are compared with simulations based on importance sampling, extending our previous results based on direct Monte Carlo simulations. We conclude by explaining why many of the standard analytical and numerical methods of large deviation theory fail in the case of anomalous large deviations.
\end{abstract}


\maketitle

\section{Introduction}

We have shown recently \cite{nickelsen2018} that time-integrated functions or observables of simple diffusions can have anomalous large deviations in the sense that their distribution decays exponentially with a scaling exponent that is nonlinear in the integration time. A simple model showing this behavior is the Ornstein--Uhlenbeck process (OUP) on $\reals$ defined by
\be
dX_t = -\gamma X_t dt +\sigma dW_t,
\ee
where $\gamma>0$ is the friction coefficient, $\sigma>0$ is the noise amplitude, and $W_t$ is a Brownian or Wiener motion representing the driving noise. Considering the integrated random variable
\be
A_T=\frac{1}{T}\int_0^T X_t^\alpha dt,
\label{obs1}
\ee
which is an estimator of the $\alpha$-moment of the stationary distribution of the OUP, we have shown that the probability density $p_T(a)$ of $A_T$ scales for integers $\alpha>2$ according to
\be
p_T(a) \sim e^{-T^\xi \bI(a)/\sigma^2}
\label{eqanoldp1}
\ee
with $\xi=2/\alpha$ in the limit of large integration time ($T\ra\infty$) and small noise amplitude ($\sigma\ra 0$) \cite{nickelsen2018}. This is to be contrasted with the scaling
\be
p_T(a) \sim e^{-T I(a)/\sigma^2}
\label{eqnormldp1}
\ee
which is usually expected to hold for random variables or observables that are integrated in time as in \eqref{obs1}.

Other processes are known to have anomalous large deviations characterised by the scaling \eqref{eqanoldp1}, including tracer dynamics \cite{krapivsky2014,sadhu2015,imamura2017}, the Kardar--Parisi--Zhang equation \cite{doussal2016,sasorov2017,corwin2018}, branching processes \cite{cox1985b,louidor2015,derrida2017}, some non-Markovian processes \cite{dembo1996b,gantert1998,zeitouni2006,harris2009}, as well as random walk models arising in queueing theory \cite{duffy2010,blanchet2013,duffy2014,bazhba2020,bazhba2022}. The simplicity of the OUP makes it a useful model for understanding anomalous large deviations with analytical methods. For this process, we find normal large deviations that scales according to \eqref{eqnormldp1} for $\alpha=1$ and $\alpha = 2$, but anomalous large deviations for $\alpha>2$ with $\xi<1$, so moments larger than 2 have fatter  tails in time. This is confirmed by mathematical results that have been reported recently for a class of diffusions that includes the OUP as a special case \cite{bazhba2022}.

In this paper, we extend these results by including the Gaussian prefactor in the instanton approximation of the path integral of $p_T(a)$, which underlies the low-noise approximation \eqref{eqanoldp1}. This prefactor, which is expressed in terms of a functional determinant, not only gives a correction to the low-noise approximation, but can also be used in the path integral to define an instanton variance, which is useful for understanding how anomalous large deviations are created in time. To test these results, we present simulations based on importance sampling that extend the direct simulations previously reported \cite{nickelsen2018}. 

The corrected $p_T(a)$ agrees remarkably well with the simulations and gives overall a good idea of the scaling of this density when considering only the long-time limit. The results on the instanton variance also support the conjecture that anomalous large deviations are created by a modified or effective process that is inherently time-dependent \cite{nickelsen2018}. By contrast, it is known that normal dynamical large deviations governed by the scaling \eqref{eqnormldp1} are created by a time-independent effective process, obtained by solving a spectral problem which happens to be ill-defined for anomalous large deviations \cite{chetrite2013,chetrite2014,chetrite2015}. 

We explore these two fluctuation mechanisms in Sec.~\ref{secres} by comparing the instanton and its variance for $\alpha=1$ and $\alpha=3$, after reviewing in Sec.~\ref{seccorr} the theory behind the instanton approximation and its Gaussian correction. We then conclude in Sec.~\ref{secconc} by explaining why many of the standard analytical and numerical techniques used in large deviation theory to study the long-time limit fail in the case of anomalous large deviations. The reasons are fundamental to the theory of large deviations and point to the need for new methods.

\section{Instanton approximation and Gaussian correction}
\label{seccorr}

Gaussian corrections to path integrals date back to the work of Gel'fand and Yaglom \cite{gelfand1960} in quantum mechanics and have been used for classical stochastic processes to derive corrections to instanton approximations of escape problems and transition pathways \cite{berglund2013,lu2017,corazza2020,ferre2021,schorlepp2021,grafke2021}, yielding temperature-dependent corrections to the original Kramer's escape result, as well as for large deviations \cite{engel2009,nickelsen2011,pietzonka2014,fatalov2014}. In this section, we present the standard approach to these corrections in which the path integral underlying $p_T(a)$ is discretised in the process space so as to perform a Gaussian integral around the instanton. In the continuum limit, the result of the Gaussian integral is expressed in terms of a functional determinant, calculated by solving a set of coupled linear differential equations with appropriate boundary conditions.  

Since the determinant depends on the instanton, we start by recalling our results \cite{nickelsen2018} about the low-noise approximation of $p_T(a)$ as well as the instanton underlying this approximation, and then presents our results for the Gaussian correction based on the functional determinant. The detailed calculations leading to the determinant are presented in the appendix.

\subsection{Instanton}

The starting point of the low-noise or instanton approximation is the path representation of $p_T(a)$:
\be
p_T(a) = \int \D[x]\, P[x]\, \delta (A_T[x]-a),
\ee
expressing this probability density as an integral over the path probability density $P[x]$ of all paths of the stochastic process leading to $A_T=a$. From the work of Onsager and Machlup \cite{onsager1953}, formalised in large deviation theory by Freidlin and Wentzell \cite{freidlin1984}, we know that $P[x]$ can be expressed, up to a normalization constant, as
\be
P[x] = e^{- S[x]/\sigma^2}
\ee
in terms of the action 
\be
S[x] = \int_0^T L(x,\dot x) dt
\ee
where 
\be
L(x,\dot x) = \frac{1}{2}(\dot x+\gamma x)^2
\ee
is the Lagrangian associated with the OUP. As a result, we can write
\be
p_T(a) =\int \D[x]\, e^{- S[x]/\sigma^2}\, \delta (A_T-a).
\label{eqPa}
\ee

This path integral is exponential with the noise amplitude $\sigma$, so it is natural to approximate it in the low-noise limit $\sigma\ra 0$ using the path having the lowest action, similarly to semi-classical approximations of quantum path integrals. The difference with the latter is that, apart from the fact that the path integral is real, we have to take into account the constraint $A_T=a$ using either a Lagrange parameter or by expressing the delta function in terms of its Laplace transform, which would add another integral in the path integral. The result of both procedures is the same: the optimal path or instanton having the lowest action, denoted by $\bx(t)$, is found by minimizing the modified action
\be
\Sa[x,\beta] = \beta T a+\int_{0}^{T} \La\big(x,\dx,\beta\big) \,d t
\label{eqSconstr}
\ee
which includes a Lagrange parameter $\beta$ dual to the constraint $A_T=a$ in the Lagrange function:
\be
\La(x,\dx,\beta) = \frac{1}{2}\big(\dx + \g x\big)^2 - \beta x^\al.
\label{eqLagr}
\ee
Equivalently, $\beta$ can be seen as the parameter of the Laplace transform that represents the delta function in the path integral. In this case, the additional Laplace integral is further approximated by a specific value of $\beta$ which is known to be equivalent to the Lagrange parameter $\beta(a)$ fixing the constraint $A_T=a$ (see \cite[App.~C.1]{touchette2009}).

The minimization of the modified action proceeds in the usual way using the Euler--Lagrange equation for the modified Lagrangian, which here takes the form
\be
\ddx(t) = \g^2 x(t) - \beta \al x(t)^{\al-1}.
\label{eqELE}
\ee
The boundary conditions are 
\be
\dx_0 - \g x_0 = 0,\qquad  \dx_T + \g x_T = 0,
\ee
since we consider open terminal conditions in which $x_0$ and $x_T$ are not a priori fixed. 

These equations can be solved analytically in the $T\ra\infty$ limit, as shown by Meerson \cite{meerson2019}, and leads to an explicit expression for the Lagrange parameter fixing the constraint $A_T=a$ \cite{nickelsen2018}:
\be
\bb(a) = \frac{1}{2} \, \g^\frac{\al+2}{\al} \, (aT)^{-\frac{\al-2}{\al}}
\left(\frac{2\sqrt{\pi}}{\al-2}\frac{\Gamma\big(\frac{\al}{\al-2}\big)}{\Gamma\big(\frac{3\al-2}{2\al-4}\big)}\right)^\frac{\al-2}{\al},
\ee
valid for $\alpha>2$. With these two results, we then find an analytical expression for $S[\bx]$, which scales with $T$ according to $T^{2/\alpha}$, yielding the scaling \eqref{eqanoldp1} for $p_T(a)$ with
\be
\tI(a) = \lim_{T\ra\infty} -\frac{\sigma^2}{T^\xi}\ln P[\bx]=\lim_{T\ra\infty} \frac{S[\bx]}{T^\xi}
\label{eqrf1}
\ee
where $\xi=2/\alpha$. The exact expression of $\bI(a)$ is too long to show (see Eq.~(13) in \cite{nickelsen2018}), but also scales like $|a|^\xi$. 

\begin{figure}
	\includegraphics[width=0.5\textwidth]{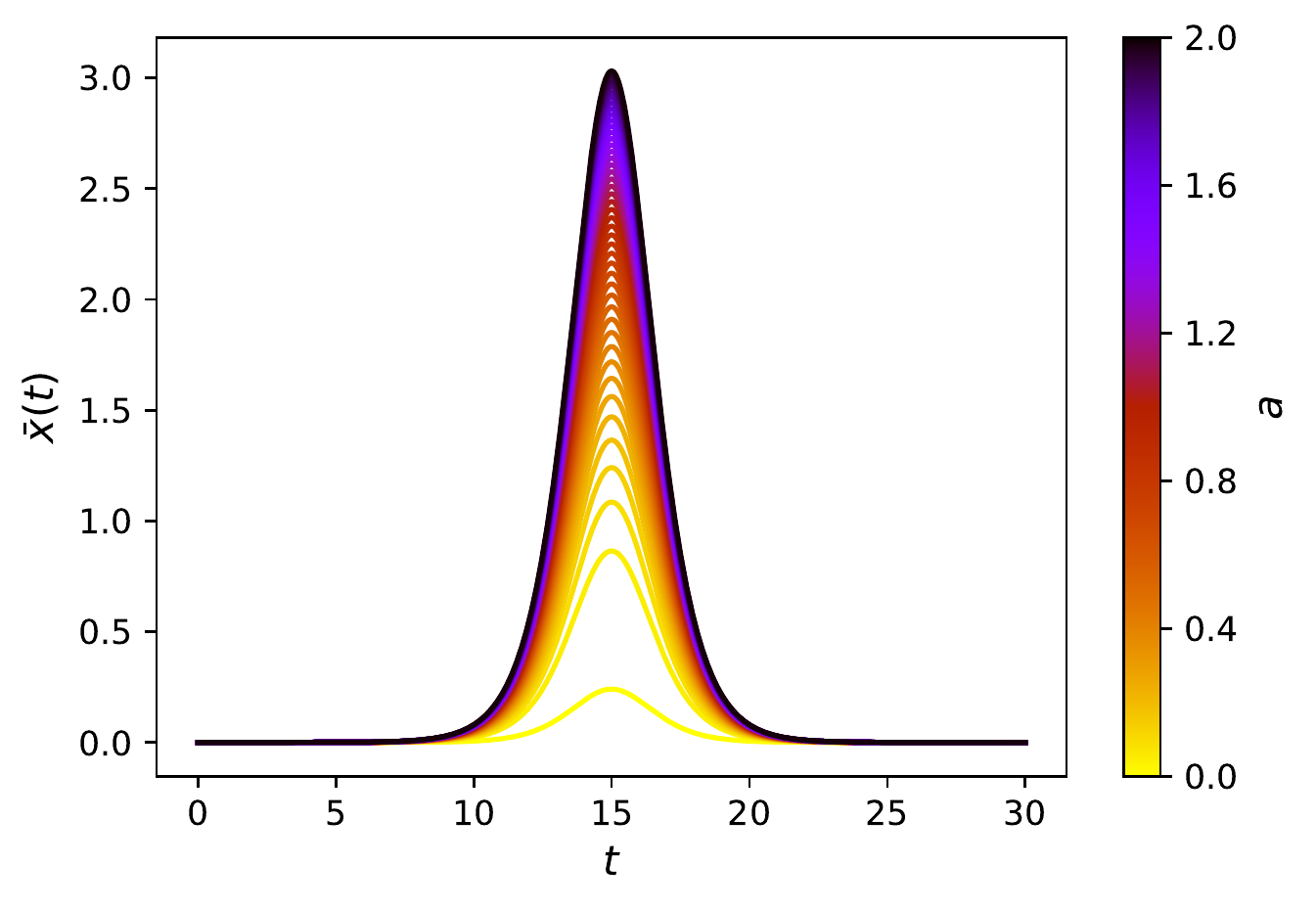}
	\caption{Instantons obtained by numerically solving the Euler--Lagrange equation \eqref{eqELE}. Parameters: $\al=3$, $\g=1$, $\s=0.5$ and $T=30$.}
	\label{figOU3xb}
\end{figure}

We recall that the instanton represents physically the path most likely to be followed (measured or observed) if we condition the process $X_t$ on the event $A_T=a$, that is, if we select only the paths of this process that realize this event. Figure~\ref{figOU3xb} shows examples of instantons for the case $\alpha=3$, obtained numerically for various values of $a$ by solving the Euler--Lagrange equation with a relaxation method \footnote{We use the SciPy function integrate.solve\textunderscore bvp to solve the Euler--Lagrange equation with boundary conditions on $x_0$ and $x_T$, and $\beta$ as an additional variable fixing the constraint.}. The time used, $T=30$, is large enough for the action to converge \cite{nickelsen2018}.

The properties of the instanton were already discussed \cite{nickelsen2018}, so we only recall the main ones needed for the results to follow when $\alpha>2$:  

(i) $\bx(t)$ has a maximum in the middle $T/2$ of the time interval simulated and is symmetric with respect to this time, attaining a value very close to $0$ for $t=0$ and $t=T$. The symmetry follows from the fact that the OUP is a reversible diffusion.

(ii) The maximum of the instanton grows with $a$ and $T$ according to
\be
\bx_{\max}(a) = \left(\frac{\gamma^2}{2\bb(a)\sigma^2}\right)^{\frac{1}{\alpha-2}}\propto (aT)^{1/\alpha}.
\ee

(iii) $\bx(t)$ is well approximated by two symmetric exponentials: one growing with rate $\gamma$ up to the maximum above, and another decaying back from this point with the same rate $\gamma$. This approximation does not capture the finite curvature of $\bx(t)$ at its maximum, but does give the correct scaling of $\tI(a)$ with $a$.

(iv) The instanton is localized over a time proportional to $1/\gamma$. This is consistent with the exponential approximation described above and explains why an integration time as short as $T=30$ reaches the large deviation limit. For longer integration times, the instanton does not change much except for its height, and its tails close to $0$ do not contribute significantly to the action for longer times. Much of the action, so to speak, happens in the localized region. 

\subsection{Gaussian correction}

The instanton solution determines the scaling
\be
p_T(a) \sim e^{-S[\bx]/\sigma^2}
\label{eqinstscaling1}
\ee
in the low-noise limit and, because of the time scaling of the action, the large deviation scaling shown in \eqref{eqanoldp1} with $\tI(a)$ as given in \eqref{eqrf1}. One way to correct this approximation is to expand the action to second-order around the instanton and to carry out the resulting Gaussian path integral so as to obtain
\be
p_T(a) \sim \frac{1}{\sqrt{D_0}} \, e^{-S[\bx]/\sigma^2} 
\label{eqpasy}
\ee
where $D_0$ is a functional determinant corresponding to continuous-time limit of the standard determinant that arises in Gaussian integrals. We refer to the scaling above with $D_0$ as the Gaussian correction of the low-noise approximation \eqref{eqinstscaling1}, which does not mean of course that $p_T(a)$ is Gaussian.

For completeness, we present the full derivation of $D_0$ in Appendix~\ref{appdet} based on the discretization of the path integral. The end result is that $D_0$ is obtained from a set of $4$ coupled linear differential equations \cite{nickelsen2011}:
\begin{align}
\ddot A(t) &= 2\g\dot A(t) - \al(\al-1)\,\bb\bx^{\al-2} A(t) \label{eqAode}\\
\dot B(t) &= \g B(t) - \frac{\al}{T}\bx^{\al-1} A(t)\label{eqBode}\\
\ddot C(t) &= 2\g\dot C(t) - \al(\al-1)\,\bb\bx^{\al-2} C(t) - \frac{2\al}{T}\bx^{\al-1}B(t)\label{eqCode}\\
\dot D(t) &= -\al(\al-1)\bb\bx^{\al-2} C(t) - \frac{2\al}{T}\bx^{\al-1} B(t) \label{eqDode}
\end{align}
with final values
\begin{align}
1 &= A(T) \nonumber\\
0 &= \dot A(T) = B(T) = C(T) = \dot C(T) = D(T).
\label{eqterm1}
\end{align}
The correction term $D_0$ corresponds to the value $D(0)$, obtained by integrating the equations above backwards in time from the terminal conditions in \eqref{eqterm1}. The solution of these equations is the main result of this paper, which we study for specific parameter values in the next section.

\subsection{Instanton variance}

The expansion of the action up to second order around the instanton can be used to define a time-dependent function, denoted by $\bv(t)$, which gives the local curvature of the path distribution $P[x]$ around the instanton and which is therefore interpreted as the variance of the path distribution along the instanton. The derivation of $\bv(t)$ is outlined in Appendix~\ref{appvar}; the result is
\be
\bv(t) = \frac{\s^2 R(t)}{D_0} A(t)A(T-t), 
\label{eqbv}
\ee
where $A(t)$ is the solution of \eqref{eqAode} and $D_0$ follows from \eqref{eqDode}. The factor $R(t)$ ensures that the constraint $A_T=a$ is met and is given by the integral
\be
R(t) = \frac{\al}{T}\bigg[ \int_0^t d \tau\, \bx(\tau)^{\al-1} r_1(\tau) + \int_t^T d\tau\, \bx(\tau)^{\al-1} r_2(\tau) \bigg],
\label{eqdefRt}
\ee
where the auxiliary functions $r_1(\tau)$ and $r_2(\tau)$ both obey the same differential equation,
\be
\ddot r(\tau) = \big(\g^2-\al(\al\!-\!1)\,\bb\bx(\tau)^{\al-2}\big)\,r(\tau) - \frac{\al}{T}\bx^{\al-1} , 
\label{eqBVPr}
\ee
but differ in their boundary conditions:
\begin{align}
0 &= \g\,r_1(0) - \dot r_1(0) \;,\quad 0 = r_1(t)\nonumber\\
0 &= r_2(t) \;,\quad 0 = \g\,r_2(T) + \dot r_2(T) , 
\label{eqBCr2}
\end{align}
which make $r_1(\tau)$ and $r_2(\tau)$ dependent on $t$. 

For large $T$, it can be shown that $R$ becomes constant,
\be
R = \frac{\al}{T} \int_0^T dt\, r(t) \, \bx(t)^{\al-1} , \label{eqdefR}
\ee
where $r(t)$ satisfies \eqref{eqBVPr} and the boundary conditions
\be
0 = \g\,r(0) - \dot r(0) \;,\quad 0 = \g\,r(T) + \dot r(T). 
\label{eqBCr}
\ee
We discuss this instanton variance and its meaning for specific parameters in the next section.

\subsection{Importance sampling simulations}

The Gaussian approximation of $p_T(a)$ shown in \eqref{eqpasy} needs to be compared with simulation results that should ideally cover a wide range of values of $A_T$. In our previous work \cite{nickelsen2018}, we  simulated large numbers of trajectories of the process $X_t$ and used them to directly estimate $p_T(a)$ and its corresponding rate function $I(a)$ by considering large simulation times \cite{touchette2011}. This method is obviously limited in that, since $p_T(a)$ is exponentially small in both $T$ and $\sigma$, an exponentially large sample is required to resolve the rate function over a wide range of values.

To improve the estimation, we can apply the idea of importance sampling by simulating a new process $Y_t$, different from $X_t$, chosen so as to make the event $A_T=a$ more likely and, ideally, to make it typical. Let $Q[x]$ denote the path distribution associated with this process. Then we can write
\be
p_T(a) =\int \D[y]\, Q[y]\, W_T[y]\, \delta (A_T-a)
\label{eqis1}
\ee
where 
\be
W_T[y] = \frac{P[y]}{Q[y]} = \exp\Big[-\frac{1}{\s^2}\big(S[y] - S_Y[y]\big)\Big],
\ee
and $S_Y[y]$ is the action of the $Y_t$ process. Thus, $p_T(a)$ can be estimated by simulating this process many times and by constructing a histogram of the samples of $A_T$ obtained, including in the histogram the likelihood factor $W_T$ computed as part of the simulation, in order to correct for the fact that we simulate $Y_t$ rather than $X_t$ \cite{touchette2011,bucklew2004,asmussen2007}. Since $Y_t$ is chosen so as to ``hit'' the event $A_T=a$ more often than $X_t$, it leads to a better estimation of $p_T(a)$ and, in turn, $I(a)$, sometimes with very few trajectories.

In practice, there are many processes that can be used to render $A_T=a$ typical. A natural one is obtained by guiding $Y_t$ along the instanton using
\be
dY_t = \dot \bx(t) dt+\sigma dW_t
\label{eqinstguide}
\ee
with $Y_0=\bx(0)$, so that, in the limit $\sigma\ra 0$, $Y(t) = \bx(t)$. This change of process has been used before in various contexts \cite{cottrell1983,dupuis1987,zuckerman1999,eijnden2012}, but was not found here to be accurate for sampling $p_T(a)$ as the noise drives trajectories far from the instanton over long times.  To mitigate this effect, we guide a linear process with the same friction as $X_t$ around the instanton using the stochastic differential equation
\be
dY_t = -\gamma (Y_t-\bx(t)) dt + \s dW_t.
\label{eqXIS}
\ee
This has the effect of producing trajectories that wander randomly around the instanton shown in Fig.~\ref{figOU3xb}. Other nonlinear friction terms were tested, but we found that the linear friction above, which defines another OUP that tracks the instanton, gives accurate results for the values of $\alpha$ considered, as it leads to a low variance for the likelihood factor $W_T$ \cite{asmussen2007}.

Note that importance sampling can be used to independently validate our theoretical results even if it uses the instanton because the estimator of $p_T(a)$ based on \eqref{eqis1} is unbiased and consistent for any modified process $Y_t$. Thus any such process can be used in principle to estimate $p_T(a)$, including the original OUP which is not guided in any way, provided that the sample is large enough. What defines a good change of process is the variance of the resulting estimator, determined by the variance of $W_T$. For more details about the efficiency of importance sampling, we refer to \cite{asmussen2007}.

\section{Results}
\label{secres}

We present in this section the results of the Gaussian correction and the instanton variance for the empirical moments of the OUP. We first consider $\alpha=1$ to test our method for normal large deviations, and then $\alpha=3$ to obtain results for anomalous large deviations. The value $\alpha=3$ is representative of all integer values $\alpha\geq 3$ leading to anomalous large deviations \cite{nickelsen2018}.

\subsection{$\alpha=1$}

For $\al=1$, all the results can be obtained exactly. For the instantons we obtain from \eqref{eqELE}
\be
\bx(t) = \frac{a\g T}{2\Omega_T^2}\,\big(2-e^{-\g t}-e^{-\g(T-t)}\big)
\ee
with
\be
\Omega_T^2 = \g T+e^{-\g T}-1.
\ee
This predicts for $T\gg 1/\gamma$ that a fluctuation $A_T=a$ is created by a constant instanton evolving close to $a$ for a time proportional to $T$. The exact action of this instanton is
\be
S[\bx] = \frac{a^2\g^3T^2}{2\Omega_T^2}
\ee
and scales like $S[\bx]\sim \gamma^2 a^2T/2$ consistently with the fact that $\bx(t)\approx a$ for $t\in [0,T]$. Consequently, $p_T(a)$ has the normal large deviation scaling \eqref{eqnormldp1} with
\be
I(a) = \frac{\gamma^2a^2}{2}.
\ee

To find the Gaussian correction to this result, we solve the coupled differential equations underlying the fluctuation determinant:
\begin{align}
A(t) &= 1 \\
B(t) &= \frac{1-e^{-\g(T-t)}}{\g T} \\
C(t) &= \frac{e^{-2\g(T-t)}-4e^{-\g(T-t)}-2\g(T-t)+3}{2\g^3T^2} \\
D(t) &= \frac{2\gamma(T-t)-2+2 e^{-\gamma(T-t)}}{T^2\gamma^2}.
\end{align}
Therefore,
\be
D_0= \frac{2\Omega_T^2}{\g^2T^2},
\ee
so that 
\be
p_T(a) \sim \sqrt{\frac{\g^2T^2}{2\Omega_T^2}} \, \exp\left[-\frac{a^2\g^3T^2}{2\s^2\Omega_T^2}\right] , 
\label{eqpaOU1}
\ee
which becomes for large $T$
\be
p_T(a) \sim \sqrt{\frac{\g T}{2}} \, \exp\left[-\frac{a^2\g^2T}{2\s^2}\right].
\label{eqdistalpha1}
\ee
The result on the right-hande side is actually the exact distribution of $A_T$ if we normalise it properly, so the instanton calculation gives in this case the correct density for all noise amplitudes. This was already noted by Onsager and Machlup \cite{onsager1953} and arises because the linear integral of a Gaussian process is also Gaussian.

\begin{figure}[t]
\includegraphics[width=0.5\textwidth]{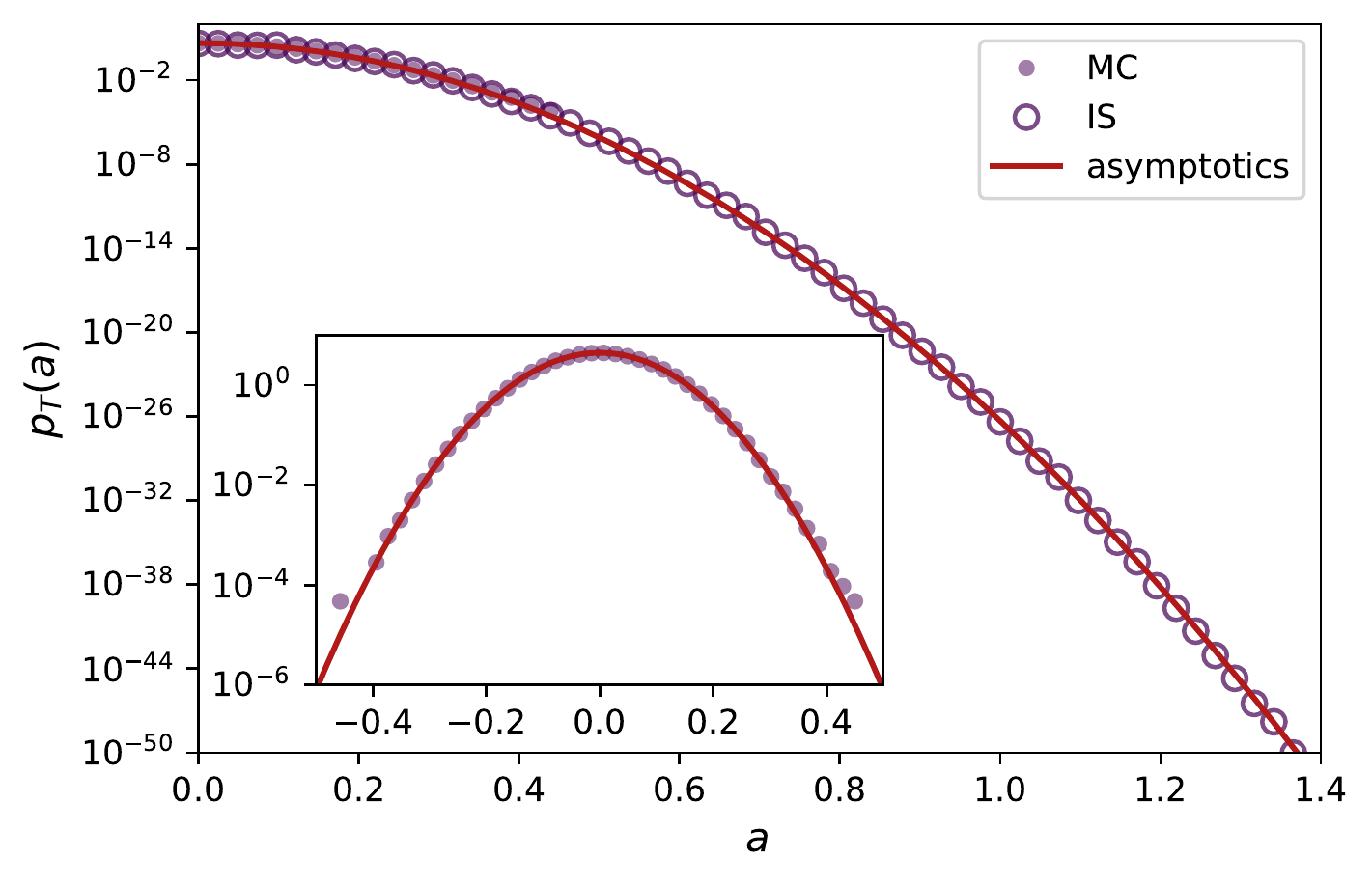}
\caption{Gaussian-corrected $p_T(a)$ (red line) compared with direct Monte Carlo (MC) simulations and importance sampling (IS) simulations. Parameters: $\g=1$, $\s=0.5$ and $T=30$.}
\label{figOU1p}
\end{figure}

We compare in Fig. \ref{figOU1p} the theoretical result \eqref{eqdistalpha1} with numerical results obtained from the direct Monte Carlo (MC) and importance sampling (IS) simulations of $p_T(a)$. The direct sampling, also shown in the inset, is naturally limited by the exponential concentration of $p_T(a)$ and the fact that this density is extremely small in the tails. Here we have used about $10^6$ sample trajectories, leading to events seen in Fig.~\ref{figOU1p} to have a density of about $10^{-6}$. The importance sampling overcomes this limitation by returning values that accurately match the theoretical distribution for values as low as $10^{-50}$ with a much smaller (and constant) sample size. The importance sampler in this case is chosen as 
\be
dY_t = -\gamma (Y_t - a)dt +\sigma dW_t
\label{eqisproc1}
\ee
to sample $A_T$ at the value $a$. This choice of dynamics, corresponding to an OUP re-centered at $a$, is known to be optimal as it has a near-constant likelihood factor in the long-time limit \cite{chetrite2015}, resulting in an estimator of $p_T(a)$ that has the least asymptotic variance \cite{guyader2020}.

This optimal property of $Y_t$ is related to the instanton variance, which can also be calculated exactly. From \eqref{eqBVPr}, we obtain the auxiliary functions
\begin{align}
r_1(\tau) &= \frac{2 - (2-e^{-\g t})e^{-\g(t-\tau)} - e^{-\g\tau}}{2\g^2T} , \\
r_2(\tau) &= \frac{2 - (2-e^{-\g(T-t)})e^{-\g(\tau-t)} - e^{-\g(T-\tau)}}{2\g^2T}
\end{align}
for the two boundary conditions in \eqref{eqBCr2}, giving
\be
R(t) = \frac{2\g T - 6 + 4(e^{-\g t}\!+\!e^{-\g(T-t)}) - (e^{-2\g t}\!+\!e^{-2\g(T-t)})}{2\g^3T^2} ,
\ee
when inserted in \eqref{eqdefRt}. From the constant $A(t)=1$, we thus find with \eqref{eqbv}
\be
\bv(t) = \frac{2\g T - 6 + 4(e^{-\g t}\!+\!e^{-\g(T-t)}) - (e^{-2\g t}\!+\!e^{-2\g(T-t)})}{4\g\Omega^2/\s^2},
\ee
which reduces to
\be
\bv(t) = \frac{\s^2}{2\g}
\ee
in the limit $T\ra\infty$. Hence, the path distribution $P[x]$ has a constant variance along the constant instanton $\bx(t)=a$, which means physically that the fluctuation $A_T=a$ can be seen as being created by a linear process with stationary mean $a$ and variance $\sigma^2/(2\gamma)$. These, as can be checked, are precisely the stationary mean and variance of the importance sampling process $Y_t$ defined above, so that this process matches the local process determined from the path distribution around the instanton.

This result is expected. From recent works \cite{chetrite2013,chetrite2014,chetrite2015}, it is known that the process $X_t$ conditioned on realising the fluctuation $A_T=a$ is equivalent in the long-time limit to another Markov process, called the effective or driven process, which happens here to be the process $Y_t$ defined in \eqref{eqisproc1} \cite[Sec.~6.2]{chetrite2014}. The construction of the driven process is known when the large deviations of $A_T$ are normal, in the sense of \eqref{eqnormldp1}, and predicts in this case that the driven process is a homogeneous process. The low-noise limit of that process gives the instanton, which explains why we obtain here a constant instanton centered at $a$ having a constant variance. 

\subsection{$\alpha=3$}

\begin{figure}[t]
\includegraphics[width=0.5\textwidth]{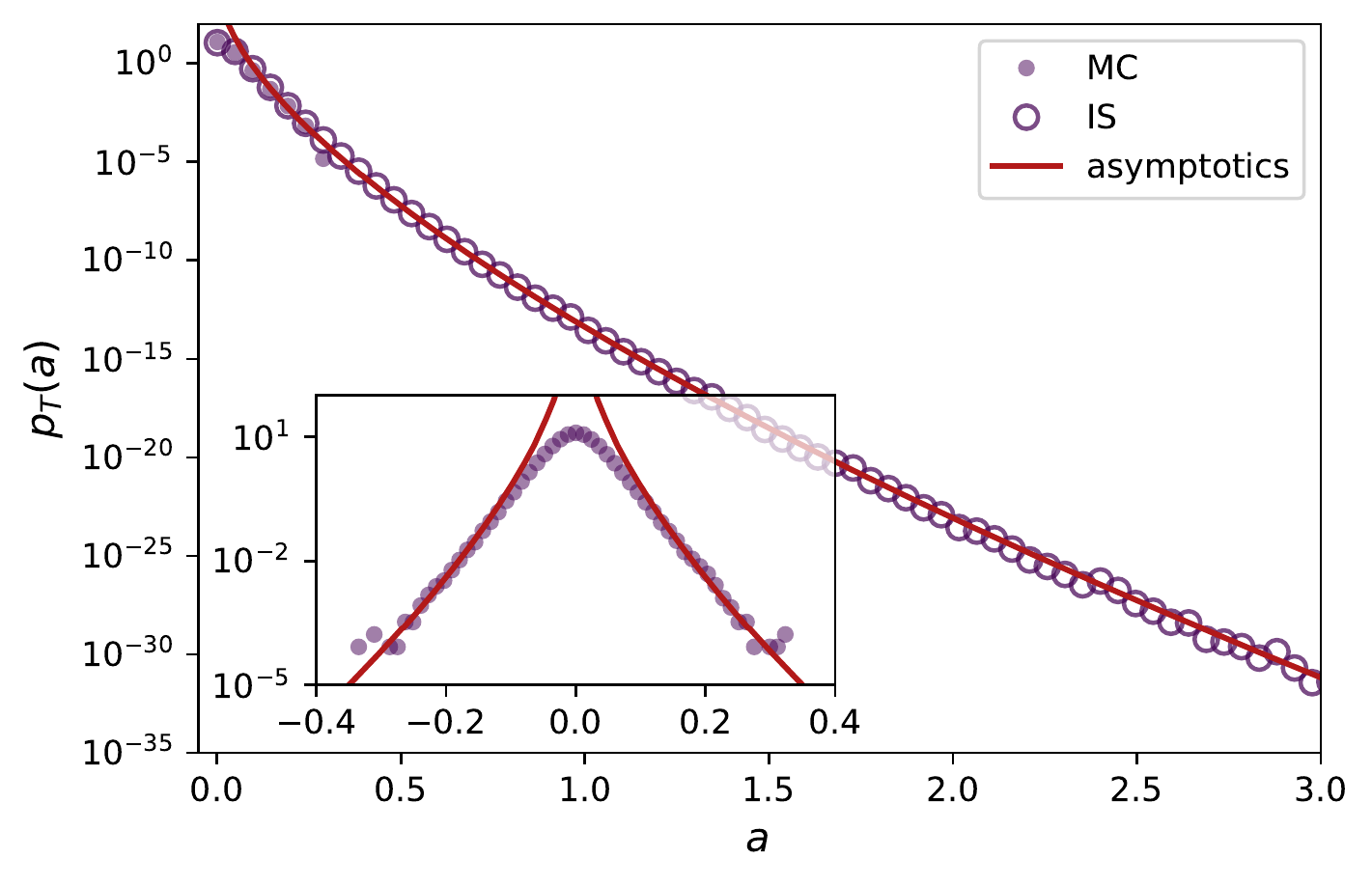}
\caption{Gaussian-corrected $p_T(a)$ (red line) for $\al=3$ compared with MC and IS simulations. Parameters: $\g=1$, $\s=0.5$ and $T=30$.}
\label{figOU3p}
\end{figure}

The instanton and fluctuation determinant cannot be found analytically for $\alpha=3$ when the integration time $T$ is finite, so we resort to obtaining them numerically. For the instanton, we solved the Euler-Lagrange equation \eqref{eqELE} with an relaxation algorithm for different $a$, using a double exponential peaked at $t=T/2$ as the initial guess. Once we have the instantons for two contiguous $a$ values, we extrapolate from these a new initial guess for the next $a$ value. We repeat this procedure until we cover a desired range of $a$ values. For the boundary solver, we use a minimal tolerance of $3\cdot10^{-14}$ and a maximum of $10^5$ mesh points. We also use $T=30$ for the integration time, which appears to be enough to give results that are in the large deviation regime \cite{nickelsen2018}. The solutions are shown again in Fig.~\ref{figOU3xb}, with the properties that we listed in the previous section, and were checked for a peak at $t=T/2$.

To obtain the fluctuation determinant, we numerically solve Eqs~\eqref{eqAode}-\eqref{eqDode}, feeding in the instantons as high resolution cubic interpolation functions and using a Runge--Kutta scheme with a minimum tolerance set to $3\cdot10^{-14}$ to control numerical instabilities. We show in Fig.~\ref{figOU3p} the approximate $p_T(a)$ obtained from \eqref{eqpasy} with the resulting value for $D_0$ as well as
\be
S[\bx] \sim T^{2/3} \bI(a)
\ee
and
\be
\tI(a) = \left(\frac{9}{10}\right)^\frac{1}{3} \gamma^\frac{5}{3} a^\frac{2}{3}.
\label{eqIOU3_zn}
\ee
These results for the action and the rate function were found in our previous study \cite{nickelsen2018}. We also show in Fig. \ref{figOU3p} the results of the MC and IS simulations based on the modified OUP \eqref{eqXIS} tracking the instanton.

\begin{figure}[t]
\centering
\includegraphics[width=0.5\textwidth]{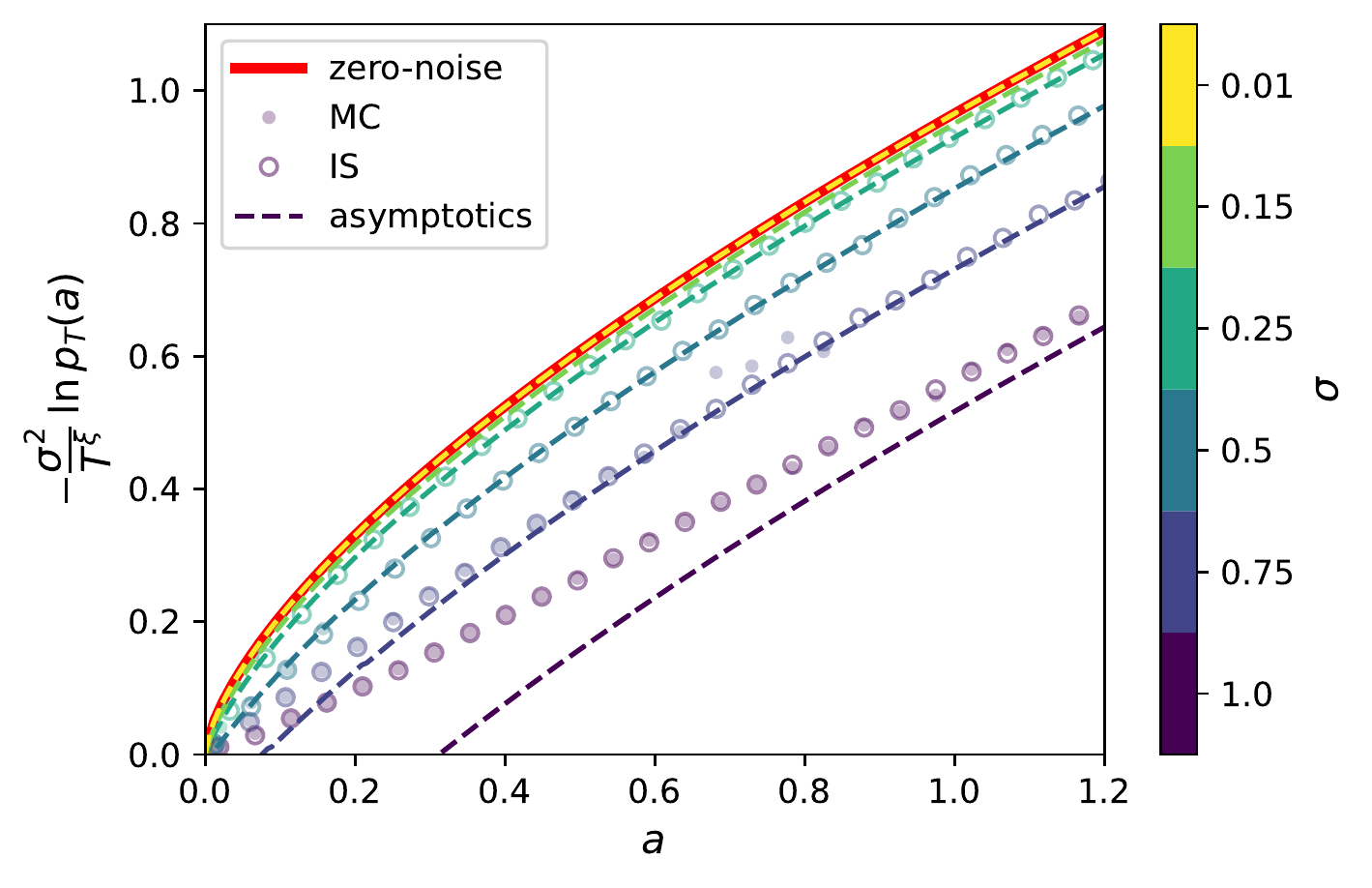}%
\caption{Log-probability scaled with $\sigma$ and $T$ for different values of $\sigma$ showing the convergence to the low-noise rate function $\tI(a)$. Parameters: $\al=3$, $\g=1$ and $T=30$.}
\label{figOU3Ia}
\end{figure}

The simulation results show again that the Gaussian correction gives a good approximation of $p_T(a)$, except now near $a=0$ where the low-noise approximation based on $\tI(a)$ above predicts a peaked maximum at $a=0$ which is actually smooth when $\sigma$ is finite. This rounding effect is illustrated in Fig.~\ref{figOU3Ia}, which shows the same data on a different scale and different values of $\sigma$. The convergence to the low-noise rate function $\tI(a)$, shown in red, and the emergence of a peak at $a=0$ are clearly seen. 

\begin{figure}[t]
\includegraphics[width=0.5\textwidth]{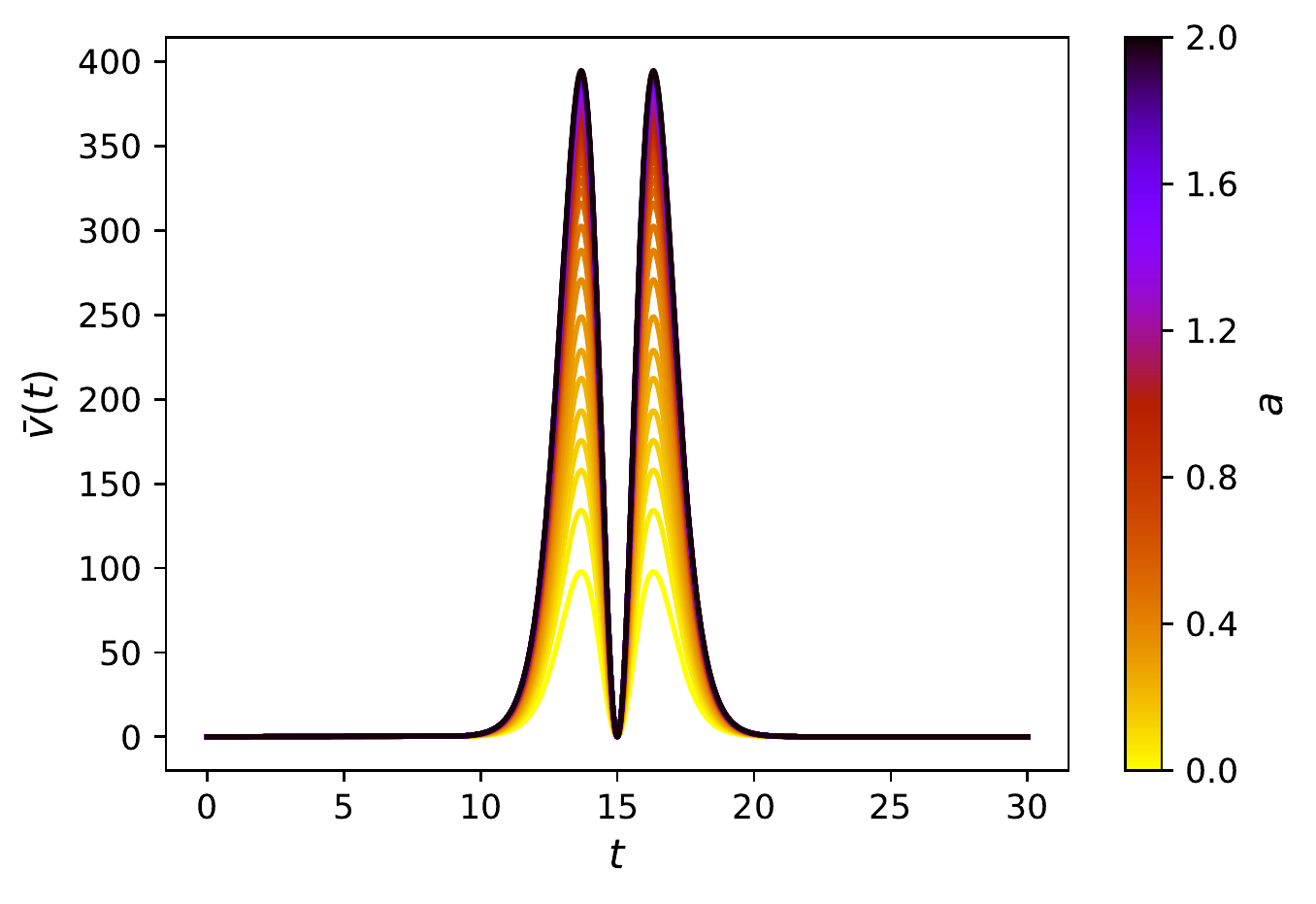}
\caption{Instanton variance $\bv(t)$ for different values of $a$. Parameters: $\alpha=3$, $\g=1$, $\s=0.5$, and $T=30$.}
\label{figOU3vb}
\end{figure}

The fact that the instanton evolves in a time-dependent way for $\alpha=3$, as seen in Fig.~\ref{figOU3xb}, makes this case very different from the case $\alpha=1$ and is what gives rise to the scaling \eqref{eqanoldp1} describing anomalous large deviations. The instanton variance $\bv(t)$ found from \eqref{eqbv} is also time-dependent, as shown in Fig.~\ref{figOU3vb}. 

Obtaining $\bv(t)$ is a challenging task, since the values involved in Eqs.~\eqref{eqBVPr} and \eqref{eqdefRt} are very small (of the order  $10^{-30}$ to $10^{-20}$). For this reason, we took care to solve these equations numerically using different mesh points and interpolations for the instanton to see if the results were stable. We found that the maximum value of $\bv(t)$, which sets the scale of the variance, cannot be relied on, since it is  sensitive to the order of approximation used for $\bx(t)$ \footnote{For some low approximations of $\bx(t)$, for example, we obtain a maximum variance of the order of $10^4$, whereas for the most accurate instanton that we have with $10^5$ points we obtain a maximum variance of 400, which is the value reported in Fig.~\ref{figOU3vb}.}, but that the double-peak shape of $\bv(t)$ seen in Fig.~\ref{figOU3vb} is stable and so is quantitatively valid. Initially, the variance is low and starts to increase when the instanton itself starts increasing to its maximum. Unlike the instanton, however, the variance has a turning point before $t=T/2$ beyond which it decreases rapidly to a low value (close to 0 from numerical calculations) precisely at $t=T/2$. After this time, the same behavior is repeated, showing overall that the fluctuations of $A_T$ are created by stochastic trajectories that follow the time-dependent instanton and fluctuate around that instanton, except at $t=T/2$, where they all converge and go through $\bx(T/2)=\bx_{\max}$ as a result of $\bv(T/2)\approx 0$.

It is difficult to verify the instanton variance independently from simulations, since it relies on rare trajectories underlying the large deviations of $A_T$ whose variance differs from the variance of the trajectories simulated with importance sampling \footnote{Estimators based on importance sampling are unbiased but generally have different variance.}. However, the behavior of $\bv(t)$ shown in Fig.~\ref{figOU3vb} agrees qualitatively with the fluctuation paths reported in our previous study \cite[Fig.~3]{nickelsen2018}, which have reduced fluctuations at the instanton peak (shifted numerically at $t=T/2$ for comparison). The fact that $\bv(t)$ is symmetric with respect to $t=T/2$ is also supported qualitatively from that figure and is consistent with the fact that the conditioning of the OUP on the event $A_T=a$ is a reversible process, since the OUP itself is reversible \cite[Sec.~5.5]{chetrite2014}. This does not mean that all stochastic trajectories realising the event $A_T=a$ have to be symmetric with respect to $t=T/2$. However, the time-reversal of any such trajectory realises the same value $A_T=a$ with the same probability, by virtue of the OUP being reversible, which means that the whole ensemble of trajectories realising $A_T=a$ must define a reversible process.

This ensemble of trajectories realising $A_T=a$ was studied extensively for normal large deviations \cite{chetrite2013,chetrite2014,chetrite2015}: it is defined mathematically as a conditioning of the path distribution $P[x]$ on the event $A_T=a$, and is known in the regime of normal large deviations to be a time-homogeneous and stationary Markov process, at least in the absence of dynamical phase transitions \cite{tsobgni2018}. The case $\alpha=1$ follows this result: the mean and variance of the instanton are time-independent, since the conditioning of $P[x]$ on $A_T=a$ is time-independent in the long-time limit. For $\alpha=3$, by contrast, we find a time-dependent instanton mean and variance, suggesting that the ensembles of paths realising $A_T=a$ is described as a whole by a time-dependent process. Similar results are found in the context of simple random walks and jump processes arising in queueing theory \cite{duffy2010,blanchet2013,duffy2014}.

Based on these results, it is natural to conjecture that a necessary condition for observing anomalous large deviations is that the ensemble of trajectories or process realising $A_T=a$ is time-dependent. In other words, if the large deviations of $A_T$ are anomalous, then the conditioned process realising those large deviations is explicitly time-dependent. This is suggested not only by our instanton results, which provide partial information about the mean and variance of that process, but also by the fact that many techniques used for obtaining large deviations do not work for anomalous large deviations, either because they assume or predict that that process is time-independent in the long-time limit. We discuss this point in more details in the next section and suggest ideas for dealing with anomalous large deviations.

\section{Concluding remarks}
\label{secconc}

In principle, the Gaussian correction of the path integral is not expected to describe $p_T(a)$ for arbitrary noise amplitudes, since it is a further approximation of the path integral in terms of $\sigma$. However, the good agreement that we find between this correction and the simulation results shows that it recovers much of $p_T(a)$, especially in the tails (see Fig.~\ref{figOU3Ia}), giving us some information about the anomalous large deviations of $A_T$ in the limit where $T\ra\infty$ with $\sigma$ finite. In this regime, $p_T(a)$ is expected to scale according to
\be
p_T(a) \sim e^{-T^\xi J(a)}
\label{eqldpanom2}
\ee
as $T\ra\infty$, so the limit function
\be
J(a) = \lim_{T\ra\infty} -\frac{1}{T^\xi}\ln p_T(a)
\ee
should exist. Mathematical estimates of this function have been reported recently for a class of diffusions that include the OUP \cite{bazhba2022}, although it is not clear whether they involve the low-noise limit. With this extra limit, the rate function that one obtains is
\be
\bI(a) = \lim_{\sigma\ra 0}\lim_{T\ra\infty} -\frac{\sigma^2}{T^\xi} \ln p_T(a) =\lim_{\sigma\ra 0} \sigma^2 J(a),
\ee
which follows from the instanton approximation, as well as logarithmic corrections of this function coming from the Gaussian prefactor \footnote{In the case $\alpha=1$, we find $D_0\sim 1/T$, so the correction is in $\ln T$ in the exponent. For $\alpha>2$, simulation results suggest a different scaling, namely, $D_0\sim T^{-2/\alpha} a^{2-2/\alpha}$, which is still logarithmic in the exponential.}. 

The reason for considering the low-noise limit, as we have argued before \cite{nickelsen2018}, is that many techniques that are standard in large deviation theory do not work in the case of anomalous large deviations. In particular, we cannot obtain $J(a)$ by applying the contraction principle to the level 2 or level 2.5 rate functions (see \cite{hoppenau2016} for details), as these all are defined in the normal scaling regime and thus predict normal large deviations for $A_T$ when the contraction has a non-trivial solution. We also cannot obtain $J(a)$ as the Legendre transform of the scaled cumulant generating function (SCGF), defined as
\be
\lambda(k) = \lim_{T\ra\infty} \frac{1}{T}\ln E[e^{TkA_T}],
\label{eqscgf1}
\ee
since the exponential in the expectation above has the wrong scaling in $T$ and, therefore, does not capture the anomalous scaling \eqref{eqldpanom2}. In fact, the SCGF, as defined above, diverges for all $k\neq 0$. This follows because the SCGF is related, in the case of normal large deviations, to the ground state energy of a quantum-like potential \cite{touchette2017}, which is not confining and has no lower bound when $\alpha>2$ and $k\neq 0$ \cite{nickelsen2018}.

To circumvent this problem, one can attempt to regularise the related quantum problem, e.g., by considering a limited range of the potential, as done recently by Smith \cite{smith2022}. This approach is able to recover the central, Gaussian part of $p_T(a)$, but seems insufficient to obtain $J(a)$, since it is based on approximating the SCGF, which is again formally divergent, and predicts a normal rather than anomalous scaling of large deviations because of the effective confinement introduced.

Another idea is to redefine the SCGF by the limit,
\be
\lambda_\xi(k)=\lim_{T\ra\infty} \frac{1}{T^\xi}\ln E[e^{T^\xi kA_T}],
\label{eqscgf2}
\ee
to match the limit \eqref{eqldpanom2} capturing the anomalous scaling of $p_T(a)$. This modified SCGF is covered by large deviation theory (see \cite[App.~D]{touchette2009}), although little is known about its properties, especially its connection with the Feynman--Kac equation and long-time solutions of this equation. In our case, $\lambda_\xi(k)$ also diverges when $\alpha>2$ because the right tail of $J(a)$, which asymptotically matches that of $\bI(a)$, is nonconvex, but there might be other processes and observables for which the modified SCGF is finite, provided that their large deviations are anomalous and have a convex rate function.

These considerations affect not only analytical methods for obtaining rate functions, but also numerical methods. For instance, the divergence of $\lambda(k)$ should be seen in runs of the cloning algorithm, since this algorithm gradually estimates the limit \eqref{eqscgf1}. In this case, one could attempt to modify the algorithm to compute the modified SCGF in \eqref{eqscgf2}, but the precise form of this modification is yet to be investigated. 

Similarly, it is not clear how importance sampling methods should be modified to account for anomalous large deviations. From our results, it seems that the appropriate way of using this method is to use a change of process that is inherently time-dependent, but finding processes that are efficient for sampling anomalous large deviations is also an open problem. The change of process used here, which has the effect of re-centering the OUP, gives good results, as we have seen, but it is not expected to be optimal in the sense that it minimizes the asymptotic variance \cite{bucklew2004}. One way to solve this problem is to include time-dependent controls in the optimal control formalism developed for normal large deviations \cite{chetrite2015}. This leads to time-dependent Hamilton--Jacobi--Bellman equations that could be solved numerically, if not analytically.

\begin{acknowledgments}
The work of D.N.\ is supported by the Oppenheimer Memorial Trust (Posdoctoral Fellowship).
\end{acknowledgments}
\appendix

\section{Calculation of fluctuation determinant}
\label{appdet}

We explain in this section how the fluctuation determinant $D_0$ is obtained in the continuous-time limit. The starting point is Laplace's method applied the finite-dimensional integral
\be
\label{eqbspspa_F}
F = \int_{\reals^N}d z_1\cdots dz_N \, e^{-\frac{1}{\s^2} f(z_1,\dots,z_N)}.
\ee
We assume that $f:\reals^N\ra \reals$ is such that the integral exists and has a unique minimum at the point $\bz = (\bz_1,\ldots,\bz_N)$ satisfying $\nabla f(\bz_1,\dots,\bz_N)=0$.
Expanding $f$ to second order around $\bz$, we obtain after carrying out the Gaussian integral,
\be
F \sim \frac{(2\pi\s^2)^{\frac{N}{2}}}{\sqrt{\det H}} \, e^{-\frac{1}{\s^2} f(\bz_1,\dots,\bz_N)} 
\label{eqLaplMeth}
\ee
in the limit $\sigma\ra 0$, where the fluctuation determinant enters as the determinant of the Hessian
\be
H_{kl} = \frac{\p^2 f(z_1,\dots,z_N)}{\p z_k \p z_l} \bigg|_{z=\bz} .
\ee
This is the Gaussian-corrected form of the Laplace approximation. Additional corrections can be obtained by considering more terms in the Taylor expansion of $f$ around $\bz$ beyond the second-order term \cite{bender1978}.

In our problem, we apply Laplace's method to the path integral representation of $p_T(a)$, given in \eqref{eqPa}, replacing the Dirac delta function by its Laplace transform \cite{engel2009,nickelsen2011}:
\be
p_T(a) = \int\frac{dx_0}{Z}\int dx_T\int\frac{dq}{2\pi \sigma^2}\int_{(0,x_0)}^{(T,x_T)} \D[x] \, e^{- \Sa[x,\beta]/\sigma^2}\, 
\ee
where $\beta = iq/T$ and
\be
\Sa[x,\beta]= \gamma x_0^2 +\beta Ta +\int_0^T \La(x,\dot x,\beta)dt,
\ee 
is the modified action. Note that, compared with \eqref{eqSconstr}, we now integrate explicitly over the final state $x_T$ and the initial state $x_0$ with the stationary density
\be
p(x_0) =\frac{e^{-\gamma x_0^2/\sigma^2}}{Z},\quad Z = \sqrt{\frac{\pi \sigma^2}{\gamma}}.
\ee
These added terms do not influence the approximation significantly, so we do not include them in the text.

Discretizing the path integral into $N$ time-slices or steps, $t_j=j\e$, $T=N\e$, $x_j=x(t_j)$, we obtain
\be
p_N(a)= \int\frac{dq}{2\pi\s^2}\int\frac{dx_0}{Z}\prod\limits_{j=1}^{N}\int \frac{dx_j}{\sqrt{2\pi\s^2\epsilon}} \;e^{-\frac{1}{\s^2} \Sa_N(x_0,\dots,x_N,\beta)} 
\label{eqPNa}
\ee
with the discretized action
\be
\Sa_N =  \gamma x_0^2+\e\sum_{j=0}^{N-1} \left[ \frac{1}{2}\left(\frac{x_{j+1}-x_j}{\e} + \g x_j\right)^2-\beta x_j^\alpha \right] + \beta a
\ee
Applying \eqref{eqLaplMeth} to the discretized path integral, we then obtain
\be
p_N(a) \sim \frac{e^{- \Sa_N(\bx_0,\dots,\bx_N,\bb)/\s^2}}{Z\sqrt{\e^N\det H}} 
\label{eqPNa_asy1}
\ee
where $H$ is the $(N+2)\times(N+2)$ Hessian with elements
\be
H_{kl} = \frac{\p^2 \Sa_N(x_0,\dots,x_N,\beta)}{\p x_k \p x_l} \bigg|_{x_j=\bx_j,\beta=\bb}
\label{eqHesS}
\ee
using $x_{N+1}=q$. Note that the exponent can also be expressed in terms of the bare action $S$, since the instanton with $\bb$ enforces the constraint $A_T=a$, so that $\Sa(\bx_0,\ldots,\bx_N,\bb) = S(\bx_0,\ldots,\bx_N)$.

We now consider the continuous-time limit by writing the Hessian as
\begin{equation}
H =
\begin{pmatrix}  
a_0/\e    & -b_0/\e     &  0       & \cdots   &  0       & i\,u_0     \\ 
-b_0/\e   &  a_1/\e     & -b_1/\e    & \cdots   &  0   & i\,u_1     \\
\vdots &         & \ddots &  \ddots     &  \vdots  & \vdots  \\
0      & \cdots  &  0      & -b_{N-1}/\e &  a_{N}/\e & i\,u_{N} \\
i\,u_0    & i\,u_1     & \cdots  &  i\,u_{N-1} &  i\,u_{N} & 0
\end{pmatrix} \label{eqH}
\end{equation}
with
\begin{align} 
a_N &= 1 , \label{eqaN}\\
a_j &= 2 - 2\e\g + \e^2\g^2 - \e^2\al(\al-1)\bb\bx_j^{\al-2} , \label{eqaj}\\
a_0 &= 1 + \e^2\g^2 - \e^2\al(\al-1)\bb\bx_0^{\al-2}  , \label{eqa0}\\
b_j &= 1 - \e\g \label{eqbj}\\
u_j &= \e\al\bx_j^{\al-1}/T , \label{equj}\\
u_N &= 0 \label{equN}, 
\end{align}
recalling that $\beta=\frac{iq}{T}$.
The matrix elements are rescaled with $\e$ such that the determinant
\begin{equation}
D_0 = - \lim_{\substack{\e\to0\\N\to\infty}}
\begin{vmatrix}  
a_0    & -b_0     &  0       & \cdots   &  0       & u_0     \\ 
-b_0   &  a_1     & -b_1     & \cdots   &  0   & u_1     \\
\vdots &         & \ddots &  \ddots     &  \vdots  & \vdots  \\
0      & \cdots  &  0      & -b_{N-1} &  a_{N} & u_{N} \\
u_0    & u_1     & \cdots  &  u_{N-1} &  u_{N} & 0
\end{vmatrix}  
\label{eqD0def}
\end{equation}
exists in the continuous limit, so that \eqref{eqPNa_asy1} becomes
\be
p_T(a)\sim \frac{e^{-S[\bx]/\sigma^2}}{Z\sqrt{D_0}}.
\ee
To arrive at this result, which differs from \eqref{eqpasy} by the added $Z$ term, we have retrieved the Hessian in \eqref{eqHesS} from the matrix in \eqref{eqD0def} by multiplying the first $N+1$ rows with $1/\e$, the last column with $\e$, and the last row and column with $i$.

To perform the continuous limit for $D_0$, we define the minor
\be
C_j =
\begin{vmatrix} 
a_{j}  & -b_{j}   &  0        & \cdots    &  0       & u_{j}    \\
-b_{j}  &  a_{j+1} & -b_{j+1} & \cdots    &  0       & u_{j+1}  \\
\vdots &          & \ddots    & \ddots    &  \vdots  & \vdots   \\
0      & \cdots   &  0        & -b_{N-1}  &  a_N     & u_N      \\
u_{j}  & u_{j+1}  &  u_{j+2}  &  u_{N-1}  &  u_N     & 0
\end{vmatrix} , \label{eqdetC}
\ee
which results from dropping the first $j$ rows and columns, and, similarly, the two auxiliary minors
\be
B_j =
\begin{vmatrix} 
-b_{j}   & 0        &  0        & \cdots   &  0       & u_{j}   \\
a_{j+1}  & -b_{j+1} &  0        & \cdots   &  0       & u_{j+1} \\
\vdots   &          & \ddots    & \ddots   &  \vdots  & \vdots  \\
0        & \cdots   & -b_{N-2} &  a_{N-1} & -b_{N-1} & u_{N-1} \\
0        & \cdots   &  0        & -b_{N-1} &  a_N     & u_N
\end{vmatrix} \label{eqdetB}
\ee
and
\be
A_j =
\begin{vmatrix} 
a_{j}    & -b_{j}   &  0       & \cdots    &  0       \\
-b_{j}   &  a_{j+1} & -b_{j+1} & \cdots    &  0       \\
\vdots   &          & \ddots   & \ddots    &  \vdots  \\
0        & \cdots   & -b_{N-2} &  a_{N-1}  & -b_{N-1} \\
0        & \cdots   &  0       &  -b_{N-1} &  a_{N}
\end{vmatrix} . \label{eqdetA}
\ee

We expand the determinant $C_j$ in the following way,
\begin{align}
C_j &= a_jC_{j+1} + b_j
\begin{vmatrix}
-b_j   & -b_{j+1} & 0        & \dots    & 0        & u_{j+1} \\
0      & a_{j+2}  & -b_{j+2} & \dots    & 0        & u_{j+2} \\
\vdots &          & \ddots   & \ddots   &          & \vdots  \\
0      & \dots    & 0        & -b_{N-1} & a_N      & u_N     \\
u_j    & u_{j+2}  & u_{j+3}  & \dots    & u_N      & 0
\end{vmatrix} \nonumber \\ 
& -(-1)^{N-j}u_j
\begin{vmatrix}
-b_j   & a_{j+1}  & -b_{j+1} & 0        & \dots    & 0        \\
0      & -b_{j+1} & a_{j+2}  & -b_{j+2} & \dots    & 0        \\
\vdots &          & \ddots   & \ddots   &          &          \\
0      & 0        & \dots    & 0        & -b_{N-1} & a_N      \\
u_j    & u_{j+1}  & \dots    & u_{N-2}  & u_{N-1}  & u_N
\end{vmatrix} \nonumber\\
&= a_jC_{j+1} + b_j\big[-b_jC_{j+2} + (-1)^{N-j}u_jB_{j+1}\big] \\ 
& \quad - (-1)^{N-j}u_j\big[-b_jB^T_{j+1}+(-1)^{N-j}u_jA_{j+1}\big]
\end{align}
to arrive at the recursion formula
\be
C_j = a_jC_{j+1} - b_j^2C_{j+2} + 2(-1)^{N-j}u_jb_jB_{j+1} - u_j^2A_{j+1} . \label{eqCj}
\ee
Similarly, we find for the two auxiliary minors the recursion formulae
\be
B_j = -b_jB_{j+1}+(-1)^{N-j}u_jA_{j+1}  \label{eqBj}
\ee
and
\be
A_j = a_j\cdot A_{j+1} - b_j^2\cdot A_{j+2} . \label{eqAj}
\ee
Together with the final conditions
\begin{align}
C_N &= u_N^2 = 0 \label{eqCN}\\
C_{N+1} &= 0 \\
B_N &= u_N = 0 \\
A_{N-1} &= a_{N-1}a_N - b_{N-1}^2 \\
A_N &= a_N, \label{eqAN}
\end{align}
we can iterate backwards to obtain the full fluctuation determinant $C_0$ in the discretized approximation.

The fluctuation determinant $D_0$ is obtained by turning the recursion formulae \eqref{eqCj}, \eqref{eqBj} and \eqref{eqAj} into differential equations. Special care must be taken to ensure convergence.

As a first step, we eliminate the alternating factor $(-1)^{N-j}$ by the replacement $B_j\mapsto (-1)^{N-j}B_j$, obtaining
\begin{align}
C_j &= a_jC_{j+1} - b_j^2C_{j+2} - 2u_jb_jB_{j+1} - u_j^2A_{j+1}\\
B_j &= b_jB_{j+1}+u_jA_{j+1} .
\end{align}
Plugging in the coefficients $a_j$, $b_j$ and $u_j$ and inspecting the order in $\e$, it turns out that $\e C_j$ is of order $\OO(1)$. Multiplying \eqref{eqCj} by $\e$ and rearranging we get 
\begin{align}
&\frac{(\e C_{j+2})-2(\e C_{j+1})+(\e C_{j})}{\e^2} = 2 \g\frac{(\e C_{j+2})-(\e C_{j+1})}{\e} \nonumber\\ 
&\quad - \al(\al-1)\bb\bx^{\al-2}_j(\e C_{j+1}) - \frac{2\al}{T}\bx^{\al-1}_jB_{j+1} + \OO(\e) . \label{eqCjdiff}
\end{align}
Similarly, we can rearrange \eqref{eqBj} and \eqref{eqAj} to obtain
\begin{equation}
\frac{B_{j+1}-B_j}{\e} = \g B_{j+1} - \frac{\al}{T}\bx^{\al-1}_jA_{j+1}
\end{equation}
and
\begin{align}
&\frac{(A_{j+2})-2(A_{j+1})+(A_{j})}{\e^2} = 2\g\frac{A_{j+2}-A_{j+1}}{\e}\\ \nonumber
&\quad -\al(\al-1)\bb\bx^{\al-2}_jA_{j+1}+\OO(\e) .
\end{align}
Taking the continuous limit, we then recover the differential equations \eqref{eqAode}, \eqref{eqBode} and \eqref{eqCode}, noted earlier, with the final conditions \eqref{eqCN}-\eqref{eqAN}.

Since the first (and last) row of $H$ deviates from the other rows, as seen in \eqref{eqH}, we have
\be
\lim_{\e\to0} \e C_0 = 0,
\ee 
and a final step is necessary to derive $D_0$. Inspecting the last recursion step from \eqref{eqCj}, we find
\begin{align}
D_0 &= \lim\limits_{\e\ra0}\left(-a_0C_1 + b_0^2C_2 + 2u_0b_0B_1 - u_0^2A_1\right) \nonumber\\ 
&= \lim\limits_{\e\ra0}\left(\frac{(\e C_2)-(\e C_1)}{\e} - 2\g \e C_2 + \OO(\e)\right) \\
&= \dot C(0) - 2\g C(0) \label{eqD0}
\end{align}
The solution for $C(t)$ can be plugged into the equation above to obtain $D_0$. From a numerical perspective, however, it is better to use $\dot D(t) = \ddot C(t) - 2\g \dot C(t)$ as in \eqref{eqDode} to avoid cancellation of small numbers involving $\dot C(0)$.

\section{Calculation of instanton variance equations}
\label{appvar}

The basis of the Gaussian correction is the Taylor expansion of the action around the instanton, which defines a multivariate Gaussian distribution in discrete time:
\begin{align}
p_N(a) &\sim e^{-\frac{1}{\s^2} S_N(\bx)} \int \prod_{i=0}^{N+1} dx_i \nonumber\\
&\quad\times \exp\Big[ -\frac{1}{2\s^2}\sum_{kl}(x_k-\bx_k)\,H_{kl}\,(x_l-\bx_l) \Big].
\end{align}
Here, we have dropped the normalisation constants and use $x_{N+1}=\beta$. From this expression, we see that the instanton given by the components $\bx_j$ in time is the mean vector of the multivariate Gaussian, while the Hessian gives the inverse covariance matrix:
\be
\varSigma = \sigma^2H^{-1} 
\label{eqdefCovMat}
\ee
describing the Gaussian fluctuations about the instanton. As a result, it is natural to define the variance of the instanton as
\be
\bv(t_k) = \varSigma_{kk}.
\label{eqdefVar_k}
\ee

To find the diagonal elements of $H^{-1}$, we use Cramer's rule
\be
(H^{-1})_{kl} = \frac{\det H_{(kl)}}{\det H} \label{eqHinv1},
\ee
where $H_{(kl)}$ denotes the matrix that results from dropping the $k$-th row and $l$-th column in $H$. Expressing this element, as defined in \eqref{eqH}, in terms of the matrix
\be
C =
\begin{pmatrix}  
a_0    & -b_0     &  0       & \cdots   &  0       & u_0     \\ 
-b_0   &  a_1     & -b_1    & \cdots   &  0   & u_1     \\
\vdots &         & \ddots &  \ddots     &  \vdots  & \vdots  \\
0      & \cdots  &  0      & -b_{N-1} &  a_{N} & u_{N} \\
u_0    & u_1     & \cdots  &  u_{N-1} &  u_{N} & 0
\end{pmatrix} 
\label{eqCmat}
\ee
which underlies \eqref{eqdetC}, we can rewrite \eqref{eqHinv1} as
\be
(H^{-1})_{kl} = \frac{\e\,\det C_{(kl)}}{\det C} 
\label{eqHinv2}.
\ee
To further simplify $\det C_{(kl)}$, we focus on the diagonal elements ($k=l$) and write in block-form
\begin{align}
\det C_{(kk)} &= \det
\begin{pmatrix} 
&\vline& & &\vline&  \\ 
A_{0,k-1} &\vline& 0 & &\vline& U_{0,k-1} \\ 
&\vline& & &\vline& \\ \hline
&\vline& & &\vline&  \\ 
0 &\vline& A_{k+1,N} & &\vline& U_{k+1,N} \\
&\vline& & &\vline&   \\ \hline
U_{0,k-1}&\vline& U_{k+1,N} & &\vline& 0 
\end{pmatrix} 
\end{align}
with
\begin{align}
A_{0,k-1} &=
\begin{pmatrix} 
a_{0}    & -b_{0}   &  0       & \cdots    &  0       \\
-b_{0}   &  a_{1} & -b_{1} & \cdots    &  0       \\
\vdots   & \ddots   & \ddots   &           & \\
0        & \cdots   & -b_{k-3} &  a_{k-2}  & -b_{k-2} \\
0        & \cdots   &  0       &  -b_{k-2} &  a_{k-1}
\end{pmatrix}, \\
A_{k+1,N} &=
\begin{pmatrix} 
a_{k+1}    & -b_{k+1}   &  0       & \cdots    &  0       \\
-b_{k+1}   &  a_{k+2} & -b_{k+2} & \cdots    &  0       \\
\vdots   & \ddots   & \ddots   &           & \\
0        & \cdots   & -b_{N-2} &  a_{N-1}  & -b_{N-1} \\
0        & \cdots   &  0       &  -b_{N-1} &  a_{N}
\end{pmatrix}, \\
U_{0,k-1} &= (u_0,\dots, u_{k-1}), \\
U_{k+1,N} &= (u_{k+1},\dots, u_N).
\end{align}

Making use of the Schur complement, we can write
\begin{align}
\det C_{(kk)} &= -\det A_{0,k-1} \cdot \det A_{k+1,N} \nonumber\\
&\qquad\times \e\,\big(U_{0,k-1}^\mr{T}\,A_{0,k-1}^{-1}\,U_{0,k-1} \nonumber\\ 
&\qquad\qquad + U_{k+1,N}^\mr{T}\,A_{k+1,N}^{-1}\,U_{k+1,N}\big) . \label{eqdetCkk}
\end{align}
Recognising $\det A_{k+1,N}$ as $A_{k+1}$ in \eqref{eqdetA}, and noting that in the continuous limit the initial conditions for the forward determinant starting at $t=0$ are the same as in \eqref{eqAN} for $t=T$, we find that
\be
\det A_{0,k-1}  \det A_{k+1,N} \to A(T-t) \cdot A(t)
\ee
for $\e\to0$.

For the quadratic form $\e\,U_{0,k-1}^\mr{T}\,A_{0,k-1}^{-1}\,U_{0,k-1}$ we define an auxiliary vector
\be
r_1 = \e\,A_{0,k-1}^{-1}\,U_{0,k-1} 
\label{eqdefr1}
\ee
and similarly define
\be
r_2 = \e\,A_{k+1,N}^{-1}\,U_{k+1,N} 
\label{eqdefr2}
\ee
for the second quadratic form. Knowing $r_1$ and $r_2$, we obtain the value of the quadratic form via the dot product
\be
R_k = U_{0,k-1}^\mr{T} \cdot r_1 + U_{k+1,N}^\mr{T} \cdot r_2 
\label{eqdefRk} .
\ee
To get $r_1$, we multiply \eqref{eqdefr1} with $\frac{1}{\e}\,A_{0,k-1}$ from the left, and obtain a linear set of equations
\be
A_{0,k-1}\,r = \e\,U_{0,k-1} , 
\label{eqLSEr1}
\ee
where we temporarily dropped the index of $r_1$. Plugging in the coefficients \eqref{eqaN}-\eqref{equN} and rearranging, we find
\be
\frac{r_{j+1}-2r_j+r_{j-1}}{\e^2} = \big(\g^2-\al(\al-1)\,\bb\bx_j^{\al-2}\big)\,r_j - \frac{\al}{T}\,\bx_j^{\al-1} 
\ee
up to terms of order $\e$, which becomes the differential equation
\be
\ddot r(\tau) = [\g^2-\al(\al-1)\,\bb\bx(\tau)^{\al-2}] r(\tau) - \frac{\al}{T} \bx(\tau)^{\al-1} 
\label{eqODEr}
\ee
in the limit $\e\to0$. This differential equation is completed by the two boundary conditions
\be
\g\,r_1(0) - \dot r_1(0) = 0 \;,\quad r_1(t) = 0 
\label{eqBCr1_app}
\ee
resulting from evaluating the first and last equations of \eqref{eqLSEr1}.

The same steps apply to $r_2$ in \eqref{eqdefr2}, leading to the same differential equation as in \eqref{eqODEr}, but with the boundary conditions 
\be
r_2(t) = 0  \;,\quad  \g\,r_2(T) + \dot r_2(T) = 0 . 
\label{eqBCr2_app}
\ee
To obtain $R(t)$, we then insert the solutions $r_1(\tau)$ and $r_2(\tau)$ into the integrals that result from taking the continuous limit of \eqref{eqdefRk}:
\be
R(t) = \frac{\al}{T}\bigg[ \int_0^t d \tau\, \bx(\tau)^{\al-1} \, r_1(\tau) + \int_t^T d\tau\, \bx(\tau)^{\al-1} \, r_2(\tau) \bigg] . 
\label{eqdefRt_app}
\ee

For large $T$, the two solutions $r_1(\tau)$ and $r_2(T-\tau)$ approach each other (as boundary values approach zero) and \eqref{eqdefRt_app} becomes	
\be
R(t) = \frac{\al}{T} \int_0^T d \tau\, \bx(\tau)^{\al-1} \, r(\tau) . 
\label{eqdefR_app}
\ee
As a result, the determinants in \eqref{eqdetCkk} become equal in the continuous limit to $- R(t) A(t)A(T-t)$. Together with $D_0$ from \eqref{eqD0} as the continuous-limit of the determinant $\det C$, we finally arrive at
\be
\bv(t) = \frac{\s^2}{D_0}R(t)A(t)A(T-t) .
\ee

\bibliography{masterbib}

\end{document}